\documentclass{article}
\usepackage{icip2023/spconf,amsmath,graphicx}

\usepackage[utf8]{inputenc}
\usepackage[pagebackref,breaklinks,colorlinks]{hyperref}
\usepackage[dvipsnames]{xcolor}
\usepackage{multirow}
\usepackage{booktabs} %
\usepackage{amssymb}
\usepackage{amsmath}
\usepackage{caption,subcaption}

\newcommand{\bs}[1]{\boldsymbol{#1}}

\title{JPEG INFORMATION REGULARIZED DEEP IMAGE PRIOR FOR DENOISING}
\name{Tsukasa Takagi$^{\star}$, Shinya Ishizaki$^{\dagger}$, Shin-ichi Maeda$^{\star}$}
\address{$^{\star}$Preferred Networks, Inc. \\ $^{\dagger}$Graduate School of Informatics, Kyoto University}

\begin{document}
\maketitle

\begin{abstract} %
Image denoising is a representative image restoration task in computer vision.
Recent progress of image denoising from only noisy images has attracted much attention.
Deep image prior (DIP) demonstrated successful image denoising from only a noisy image by inductive bias of convolutional neural network architectures without any pre-training.
The major challenge of DIP based image denoising is that DIP would completely recover the original noisy image unless applying early stopping.
For early stopping without a ground-truth clean image, we propose to monitor JPEG file size of the recovered image during optimization as a proxy metric of noise levels in the recovered image.
Our experiments show that the compressed image file size works as an effective metric for early stopping.
\let\thefootnote\relax\footnote{Copyright 2023 IEEE. Published in 2023 IEEE International Conference on Image Processing (ICIP), scheduled for 8-11 October 2023 in Kuala Lumpur, Malaysia. Personal use of this material is permitted. However, permission to reprint/republish this material for advertising or promotional purposes or for creating new collective works for resale or redistribution to servers or lists, or to reuse any copyrighted component of this work in other works, must be obtained from the IEEE. Contact: Manager, Copyrights and Permissions / IEEE Service Center / 445 Hoes Lane / P.O. Box 1331 / Piscataway, NJ 08855-1331, USA. Telephone: + Intl. 908-562-3966.}
\end{abstract}


\section{Introduction} %
\label{sec:introduction}

Image denoising is one of the important tasks in computer vision.
Recent attractive image denoising methods tackle the case in which only noisy images are given.
In this setting, we cannot apply supervised learning which usually needs a lot of pairs of clean and noisy images for training. %

Deep image prior (DIP)~\cite{dip-ulyanov-2018-cvpr,dip-ulyanov-2020-ijcv} is an image denoising method applicable to this setting.
The key idea of DIP is to make use of the implicit regularization brought by the convolutional neural network architectures.
DIP could estimate a clean image only from a single noisy image without training the network beforehand. In DIP, the network parameters are just randomly initialized and optimized so as to reconstruct the given noisy image.
To prevent reconstructing the original noisy image, 
several follow-up DIP works have proposed to incorporate systematic early-stopping (ES)~\cite{self-validation-li-2021-bmvc,es-wmv-anonymous-2023-iclr-review}.
However, existing ES methods struggle to compute a stable metric because there is no absolute aesthetic criterion in denoising.
Also, feasible denoising during optimization depends on the noise level and the image content.
For example, the best result could be that noises can remain in the DIP-based denoised image at some high noise level.

\begin{figure}[t]
    \centering
    \begin{subfigure}[t]{0.115\textwidth}
        \centering
        \includegraphics[width=1.0\linewidth]{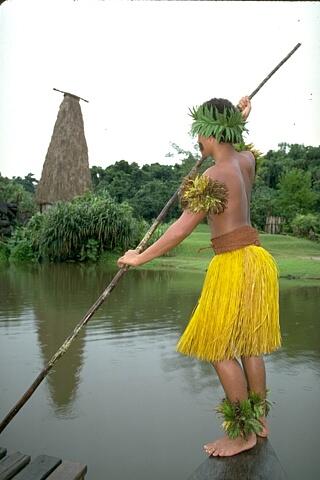}
        \vspace{-1.5em}
        \caption{$\sigma$ = 0}
    \end{subfigure}
    \hfill
    \begin{subfigure}[t]{0.115\textwidth}
        \includegraphics[width=1.0\linewidth]{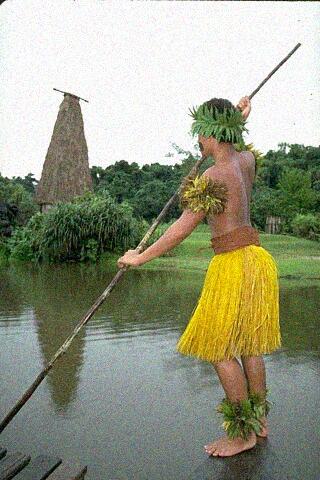}        
        \vspace{-1.5em}
        \caption{$\sigma$ = 15}
    \end{subfigure}
    \hfill
    \begin{subfigure}[t]{0.115\textwidth}
        \includegraphics[width=1.0\linewidth]{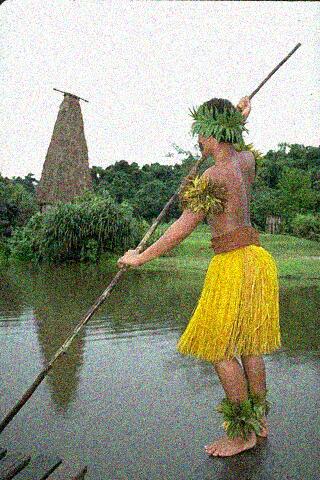}        
        \vspace{-1.5em}
        \caption{$\sigma$ = 25}
    \end{subfigure}
    \hfill
    \begin{subfigure}[t]{0.115\textwidth}
        \includegraphics[width=1.0\linewidth]{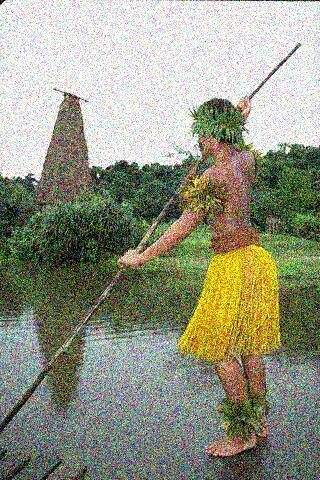}        
        \vspace{-1.5em}
        \caption{$\sigma$ = 50}
    \end{subfigure}

    \hfill
    \vspace{-0.4em}
    \begin{subfigure}[t]{0.48\textwidth}
        \centering
        \includegraphics[width=1.0\linewidth]{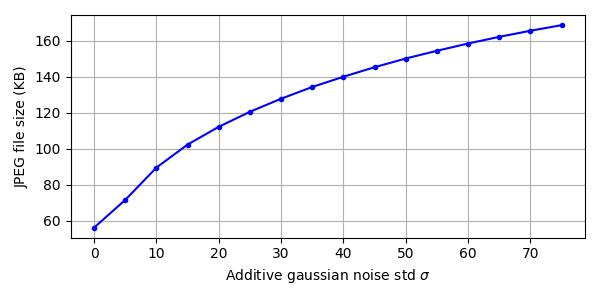}
    \end{subfigure}
    \vspace{-2.4em}
    \caption{\textbf{Relationship between different additive gaussian noise levels and JPEG file sizes.} The JPEG file sizes increase according to the noise levels in the images.}
    \label{fig:jpg-file-size}
\end{figure}

In this study, we propose a simple, but effective heuristic to determine the ES.
We propose to use compressed image file size (CIFS), in particular, JPEG compressed file size  of the reconstructed image to determine the termination.
Since the target of the image compression is mainly clean images, JPEG file size tends to increase as the noise level increases.
We preliminarily examined the relationship between the additive gaussian noise levels and their JPEG file sizes as shown in Fig. \ref{fig:jpg-file-size}.
As we can see, the JPEG file size monotonically increases according to the noise level increases.
Our ES utilizes this relationship as the proxy indicator of the reconstructed noise in the denoised image.
In our ES criterion, the optimization should be stopped where the CIFS increases although the image content would be reconstructed.

\section{Related Work} %
\label{sec:related-work}

Since image denoising is one of the long standing problems in computer vision, there are many related works.
For the sake of relevance to this paper, we overview only deep image prior (DIP) and its subsequent works employing ES to prevent overfitting.

Deep image prior (DIP)~\cite{dip-ulyanov-2018-cvpr,dip-ulyanov-2020-ijcv} demonstrates that randomly initialized neural networks can be used as an image prior for standard inverse problems including denoising.
Although DIP can optimize the network parameters when only given a noisy image in denoising, the performance would degrade because of overfitting to the noisy image.
One of the typical approaches to avoid overfitting is ES, in which some criterion is used as a proxy to measure the difference between the network output and the unknown clean image.

Several subsequent ES works of DIP have been proposed.
DIP-SURE~\cite{dip-sure-metzler-2019-baspw} introduce Stein's unbiased risk estimator (SURE) to compute approximately unbiased estimate for gaussian noise.
DIP-denoising~\cite{dip-denoising-jo-2021-iccv} extends DIP-SURE for poisson noise.
Self-validation~\cite{self-validation-li-2021-bmvc} and ES-WMV~\cite{es-wmv-anonymous-2023-iclr-review} are different directional ES methods which do not assume noise types and levels in a noisy image.
Self-validation trains an autoencoder from windowed consecutive reconstructed images and the autoencoder evaluates the next reconstructed image quality as the ES criterion.
ES-WMV computes windowed moving variance (WMV) as the ES criterion.

Our proposed ES method also does not assume noise types and levels.
Our ES criterion is based on JPEG compressed image file size (CIFS).
\section{Proposed Method} %
\label{sec:proposed-method}

\subsection{Preliminaries: Deep Image Prior (DIP)}

Before describing our proposed ES method, we describe deep image prior (DIP). DIP is proposed for image restoration problems including denoising. Image restoration algorithms aim to recover an unknown clean image $\bs{x}$ given a corrupted image $\bs{x}_0$. DIP argued that a neural network architecture behaves as the general image prior, and the clean image $\bs{x}$ can be recovered from the corrupted image $\bs{x}_0$ without additional regularizations:
\begin{equation}
    \label{eq:objective}
    \min_{\bs{\theta}} \mathcal{L} (f_{\bs{\theta}} (\bs{z}), \bs{x}_0),
\end{equation}
where $\bs{z}$ is randomly initialized and fixed noise, $f_{\bs{\theta}}$ is the neural network parameterized by $\bs{\theta}$, and $\mathcal{L}$ is the loss function such as mean squared error.

\subsection{Compressed Image File Size based ES}
\label{ss:cifs}

As described in Sec.~\ref{sec:introduction}, compressed image file size (CIFS) can be used as a proxy metric of degrees of noise.
As the network parameters fit to the corrupted image, the loss function $\mathcal{L}$ decreases almost consistently. On the contrary, the network gradually outputs a noisy image which would have a larger compressed image file size.
Based on this insight, our ES considers finding trade-off between the loss function $\mathcal{L}$ and CIFS based regularizer $\mathcal{R}$ formulated as:
\begin{equation}
    \label{eq:es-criterion}
    E(\lambda, t; \bs{z}) := \lambda \mathcal{L} (f_{\bs{\theta}_t} (\bs{z}), \bs{x}_0) + \mathcal{R}(C(f_{\bs{\theta}_t} (\bs{z}))),
\end{equation}
where ${C}$ is a function which takes an image as input and outputs CIFS, $t \in \mathbb{N}$ is an epoch number during optimization, and $\lambda \in \mathbb{R}_{\geq 0}$ is a balancing weight between two terms.
In our experiments, we adopt mean squared error as the loss function $\mathcal{L}$ same as DIP, JPEG file size output function as $C$, and squared JPEG file size averaged over image size as $\mathcal{R}$.
Specifically, $\mathcal{R}$ is
\begin{equation}
    \label{eq:regularizer}
    \mathcal{R}(L) = \frac{L^2}{HW},
\end{equation}
where $L$ is CIFS and $(H, W)$ are image height and width, respectively. Since CIFS depends on its image size, $L$ is averaged over its image size $HW$. Moreover, we also use squared file size $L^2$ so that regularization works strongly as the file size increases.

\begin{figure}[t]
    \centering
    \includegraphics[width=0.45\textwidth]{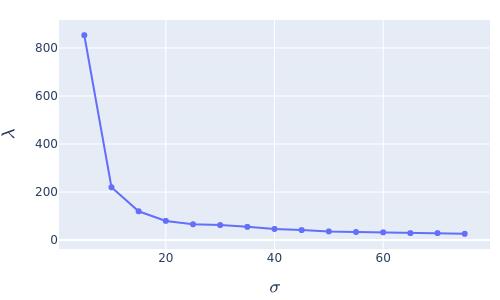}
    \vspace{-0.9em}
    \caption{\textbf{The gaussian noise standard deviation $\sigma$ and the estimated $\lambda$ in Eq.~(\ref{eq:es-criterion}).}}
    \label{fig:std-vs-lambda}
\end{figure}

The $\lambda$ is affected by degrees of noise.
We show estimated $\lambda$ for additive gaussian noise with different standard deviation $\sigma$ in Fig.~\ref{fig:std-vs-lambda}.
The $\lambda$ estimation is based on our preliminary experiment on CBSD500~\cite{cbsd500}.
We assume that we can observe 400 clean images on CBSD500.
Under this assumption, we could search $\lambda$ by the following steps: (1) adding a gaussian noise to a clean image to create a noisy image synthetically, (2) denoising the noisy image by DIP and monitoring values $\mathcal{L}, \mathcal{R}$ in Eq.~(\ref{eq:es-criterion}), (3) and finding the optimized $\lambda$ as maximizing the average PSNR between the clean image and the denoised image at the ES criterion minimized epoch over the observed 400 images.
Specifically, the final step is formulated as:
\begin{align*}
    \max_{\lambda} \mathbb{E}_{ \bs{z} \sim U_z(\bs{z}), \bs{x} \sim U_x(\bs{x}) } & 
    [ PSNR ( \bs{x}, f_{\bs{\theta}_{\hat{t}}} (\bs{z}) ) ] \\
    & \mathrm{\ where\ } \hat{t} = \underset{t}{\arg\min}  E (\lambda, t; \bs{z}),
\end{align*}
where $U_{\bs{z}}$ is a pixel-wise uniform distribution on $[0, 1]$, and $U_{\bs{x}}$ is a uniform distribution on the 400 observed images of CBSD500.

\section{Experiments} %
\label{seq:experiments}

\begin{figure}[h]
    \centering
    \includegraphics[width=0.5\textwidth]{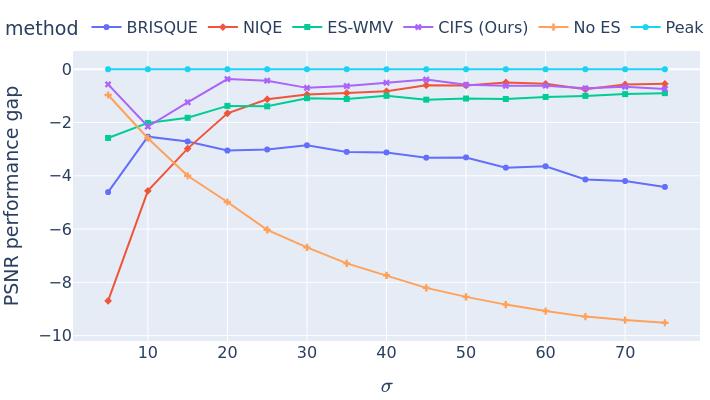}
    \vspace{-2.25em}
    \caption{\textbf{PSNR performance gap comparison on CBSD68~\cite{cbsd68} for additive gaussian noise with different standard deviation $\sigma$.}}
    \label{fig:psnr-perf-gap}
\end{figure}

\begin{table*}[h]
\centering
\caption{\textbf{PSNR comparison on CBSD68~\cite{cbsd68} and Kodak24~\cite{kodak24}.} 
Each value in the table denotes  mean $\pm$ std of PSNR for the corresponding method and dataset. 
$\sigma$ denotes the std of the gaussian noise.
"No ES" indicates the PSNR at the last epoch ($T=20$k epoch) and "Peak" indicates the maximum PSNR between a clean image and a denoised image in  $T=20$k epochs. %
}
\label{tab:quantitative-comparison}
\resizebox{1.0\textwidth}{!}{
\begin{tabular}{cccccc|cc}
\toprule
Dataset & $\sigma$ & BRISQUE~\cite{brisque-2012-tip} & NIQE~\cite{niqe-2013-spl} & ES-WMV~\cite{es-wmv-anonymous-2023-iclr-review} & CIFS (Ours) & No ES & Peak \\
\midrule
\multirow{3}{*}{CBSD68~\cite{cbsd68}}
& 15 & 27.750$\pm$2.108 & 27.484$\pm$4.172 & 28.640$\pm$3.877 & \textbf{29.223}$\pm$\textbf{2.618} & 26.464$\pm$0.535 & 30.463$\pm$2.066 \\
& 25 & 24.754$\pm$2.740 & 26.642$\pm$2.759 & 26.375$\pm$3.577 & \textbf{27.338}$\pm$\textbf{2.634} & 21.740$\pm$0.410 & 27.767$\pm$2.206 \\
& 50 & 21.078$\pm$2.183 & 23.782$\pm$2.369 & 23.289$\pm$2.826 & \textbf{23.803}$\pm$\textbf{2.321} & 15.843$\pm$0.416 & 24.388$\pm$2.241 \\
\cmidrule(lr){1-2} \cmidrule(lr){3-8}
\multirow{3}{*}{Kodak24~\cite{kodak24}}
& 15 & 29.685$\pm$2.211 & 29.247$\pm$3.768                   & 29.779$\pm$3.905 & \textbf{31.263}$\pm$\textbf{1.771} & 28.567$\pm$0.533 & 31.585$\pm$1.730 \\
& 25 & 26.341$\pm$4.906 & 27.498$\pm$3.490                   & 27.686$\pm$3.272 & \textbf{28.804}$\pm$\textbf{1.947} & 24.188$\pm$1.651 & 29.040$\pm$1.853 \\
& 50 & 22.759$\pm$3.426 & \textbf{25.224}$\pm$\textbf{1.807} & 24.640$\pm$2.322 & 25.197$\pm$1.759                   & 17.385$\pm$0.328 & 25.668$\pm$1.844 \\
\bottomrule
\end{tabular}
} %
\end{table*}

\begin{figure*}[t]

\centering
\begin{subfigure}[h]{\textwidth}
    \centering
    \includegraphics[width=\textwidth]{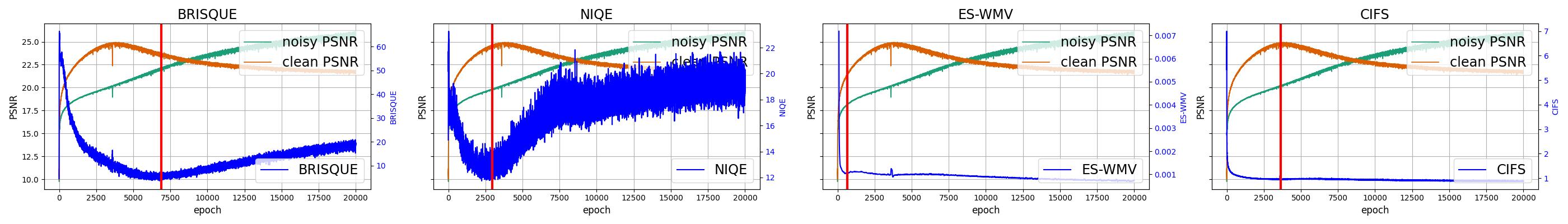}
\end{subfigure}
\hfill
\vspace{-4pt}
\centering
\begin{subfigure}[h]{\textwidth}
    \centering
    \includegraphics[width=\textwidth]{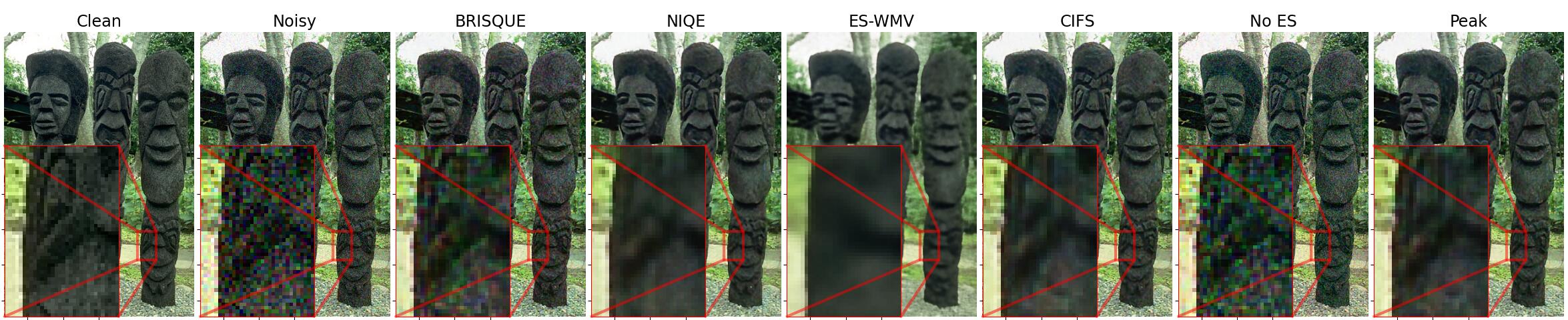}
\end{subfigure}
\hfill
\vspace{-2pt}

\centering
\begin{subfigure}[h]{\textwidth}    
  \centering
  \includegraphics[width=\textwidth]{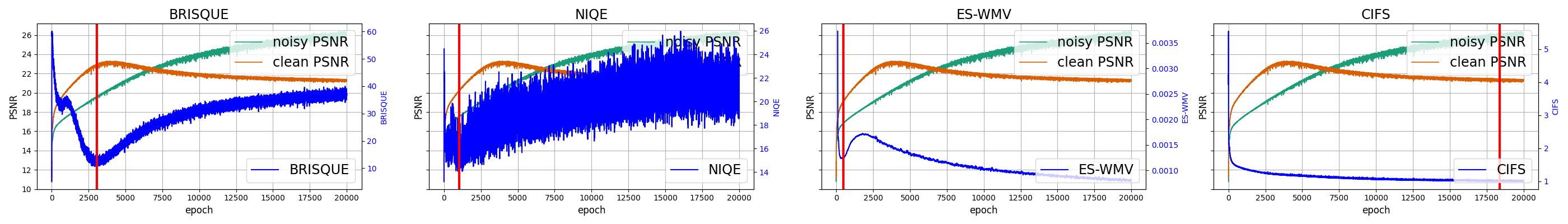}
\end{subfigure}
\hfill
\vspace{-4pt}
\centering
\begin{subfigure}[h]{\textwidth}    
  \centering
  \includegraphics[width=\textwidth]{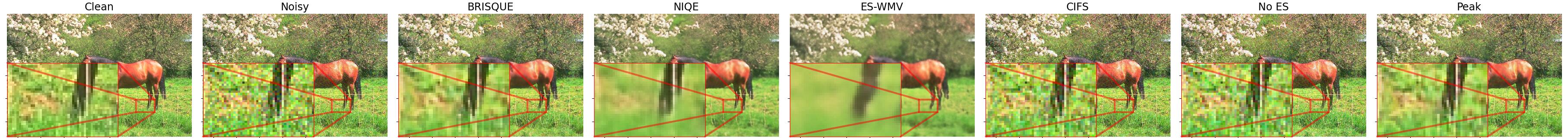}
\end{subfigure}
\hfill
\vspace{-2pt}

\centering
\begin{subfigure}[h]{\textwidth}
  \centering
  \includegraphics[width=\textwidth]{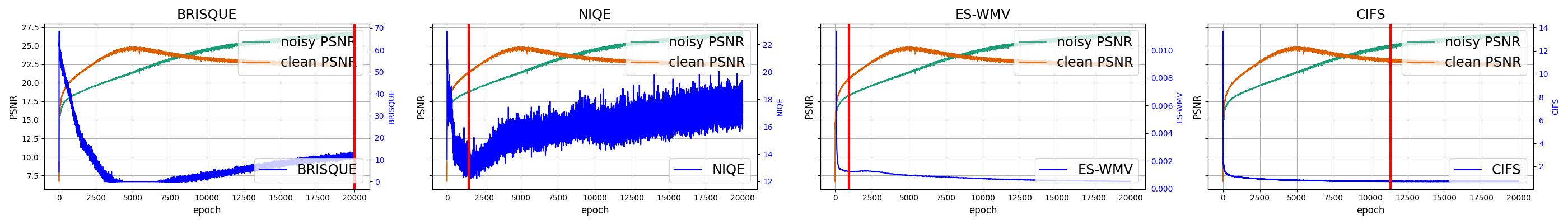}
\end{subfigure}
\hfill
\vspace{-4pt}
\centering
\begin{subfigure}[h]{\textwidth}
  \centering
  \includegraphics[width=\textwidth]{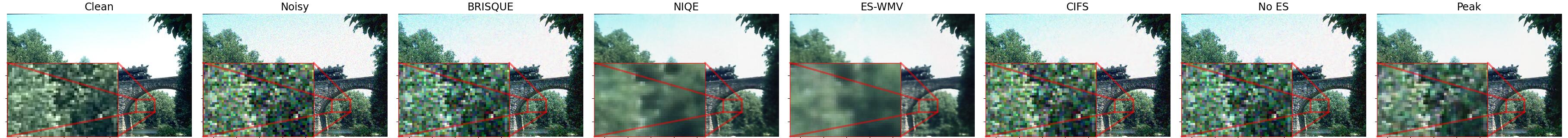}
\end{subfigure}
\hfill
\vspace{-2pt}

\vspace{-4.5pt}
\caption{\textbf{Qualitative comparison for $\sigma = 25$ on CBSD68~\cite{cbsd68}.} CIFS almost always found good ES epochs whereas the comparative methods sometimes failed to find good ES epochs.}
\label{fig:qualitative-comparison}
\end{figure*}

\subsection{Datasets}

We adopted CBSD68~\cite{cbsd68} and Kodak24~\cite{kodak24} as image denoising benchmark datasets.
CBSD68 is a different split not overlapping the split used in $\lambda$ estimation described in Sec.~\ref{ss:cifs}.

\subsection{Evaluation process}

Non-reference image quality assessment (NR-IQA) has tackled to score image quality without a reference image.
Since NR-IQA methods are expected to provide good image criteria, these methods can be employed as ES.
Following the recent ES method for DIP, ES-WMV~\cite{es-wmv-anonymous-2023-iclr-review}, we compared to BRISQUE~\cite{brisque-2012-tip} and NIQE~\cite{niqe-2013-spl} as the NR-IQA methods and ES-WMV~\cite{es-wmv-anonymous-2023-iclr-review} as the ES method to validate whether our proposed method can provide a good ES criterion. We used models and code provided by OpenCV\footnote{https://github.com/opencv/opencv\_contrib}~\cite{opencv_library} and LIVE\footnote{https://github.com/utlive/live\_python\_qa} for BRISQUE and NIQE, respectively.

Let $T$ denote the number of training epochs.
First, we optimized the neural network parameters with respect to Eq.~(\ref{eq:objective}) reaching to $T$ epochs.
Second, we obtained the ES detected epoch $t_{\ast}$ such that each ES criterion is satisfied.
Following ES-WMV ES detection, our and other ES criteria find ES candidate epochs $\mathcal{T}$ and adopt the minimum epoch in $\mathcal{T}$ as the ES detected epoch $t_{\ast}$.
The candidate epoch $t \in \mathcal{T}$ satisfies that the ES criterion outputs a smaller value than the next consecutive $S$ epochs.
Specifically,
\begin{equation}
\mathcal{T} = \{ t\ |\ t = \underset{\tau \in [t, t + S]}{\arg\min} M \left( f_{\bs{\theta}_{\tau}} (\bs{z}) \right) \},
\end{equation}
where $M$ is a specific criterion function for each ES method.
If an ES method cannot find any candidate epochs (i.e., $\mathcal{T} = \emptyset$), $T$ is adopted as an alternative to the ES detected epoch.
Finally, PSNR is computed between the clean image $\bs{x}$ and the denoised image $f_{\bs{\theta}_{t_{\ast}}} (\bs{z})$ as the ES criterion evaluation.

\subsection{Implementation details}

CIFS utilizes JPEG as image compression method. JPEG can specify a quality value $Q \in [0, 100]$ (higher is better image quality) which affects the saved image file size. $Q$ is set to 95 which is the default value in OpenCV.

The following other implementation details are shared with CIFS and all comparative methods.
Pixel-wise gaussian noise is independently sampled from $\mathcal{N}(0, \sigma^2)$ where we set $\sigma \in \{5, 10, \ldots, 75 \}$ for pixel value range $[0, 255]$.
The network architecture is same as DIP and Adam~\cite{adam-2015-iclr-kingma14} with a learning rate $0.01$ is used.
The input noise $ \bs{z} \in \mathbb{R}^{D \times H \times W} $ to the network is randomly initialized and fixed during optimization where $D$ is set to $32$.
We adopted the $\bs{z}$ perturbation technique from DIP~\cite{dip-ulyanov-2018-cvpr,dip-ulyanov-2020-ijcv} where we perturb $\bs{z}$ at each iteration.
The number of training epochs $T$ is set to 20k.
$S$ is set to 1k, which is same as ES-WMV experimental setting.

\subsection{Analysis}

Table~\ref{tab:quantitative-comparison} shows PSNR comparisons of our and other comparative ES methods for $\sigma \in \{15, 25, 50\}$ on CBSD68 and Kodak24.
Fig.~\ref{fig:psnr-perf-gap} depicts PSNR performance gap comparison for $\sigma \in \{5, 10, \ldots, 75\}$ on CBSD68 as the broader noise level range evaluation.
DIP without ES ("No ES") is significantly lower than the maximum achievable PSNR ("Peak") regardless of gaussian noise levels.
CIFS provides good ES criterion and almost outperforms other comparative methods for both datasets.
Moreover, the standard deviations of PSNR for different noise levels are relatively small compared to other ES methods.
ES criteria and qualitative comparison for $\sigma = 25$ are depicted in Fig.~\ref{fig:qualitative-comparison}.
Note the "Peak" PSNR is same for all ES methods since all ES metrics are computed in the same optimization process.
BRISQUE and NIQE, which are NR-IQA methods, have an issue with the magnitude of variance of their metrics.
ES-WMV is stable, however, it keeps a long variance sequence since the metric is windowed moving variance.
CIFS can compute the metric independently at each iteration and still provide a stable metric.

\section{Conclusion} %
\label{seq:conclusion}

We propose a novel ES for DIP designed for image denoising.
Our ES employs JPEG compressed image file size as a proxy metric to measure degrees of noise for denoised images.
Our method provides a good ES criterion and outperforms many of the other comparative methods.
Moreover, our ES method can compute the metric independently at each iteration and still provide a stable metric.

\subsection*{Acknowledgement} %

We thank Sol Cummings for helpful feedback.
This research work was financially supported by the Ministry of Internal Affairs and Communications of Japan with a scheme of "Research and development of advanced technologies for a user-adaptive remote sensing data platform" (JPMI00316).

\clearpage
\bibliographystyle{icip2023/IEEEbib}
\bibliography{refs/esdip}

\end{document}